\documentstyle[12pt]{article}
\begin{document}
\pagestyle{plain}

\begin{titlepage}
\hfill Preprint LITP-94/CP-03 and hep-lat/9404011
\begin{center}
\vspace*{3cm}
{\Large \bf
             Critical region of the random bond Ising model
}\\[4ex]

{\bf A.L. Talapov\dag\S\ and L.N. Shchur\dag}
\\[1ex] {\sl \dag\ Landau Institute for Theoretical Physics,\\ GSP-1 117940
Moscow V-334, Russia\\
\S\ Physikalisches Institut der Universit\"{a}t Bonn\\
Nu\ss allee 12, 53115 Bonn, Germany}
\end{center}


\vskip 4ex

\begin{center}
{\bf Abstract}\\[1.5ex]
\end{center}
\indent
We describe results of the cluster algorithm Special Purpose Processor
simulations of the 2D Ising model with impurity bonds. Use of large
lattices, with the number of spins up to $10^6$, permitted to define
critical region of temperatures, where both finite size corrections
and corrections to scaling are small. High accuracy data
unambiguously show increase of magnetization and magnetic susceptibility
effective exponents $\beta$ and $\gamma$,
caused by impurities. The $M$ and $\chi$ singularities
became more sharp, while the specific heat singularity is smoothed.
The specific heat is found to be in a good agreement with Dotsenko-Dotsenko
theoretical predictions in the whole critical range of temperatures.
\end{titlepage}


\section{Introduction}

The problem of inhomogenities influence on phase transitions has
a long history.
The first model with some kind of impurities to be studied,
is, due to its simplicity, the Ising model.
Effect of randomness on the critical behaviour of different Ising models has
been investigated by Harris \cite{Harris}.
He found that if the specific heat of the
pure system diverges as some power of $(T_c-T)^{-1}$,
then the critical behaviour
will be changed by impurities. Such a result does not give any information
for the 2D Ising model, which has a logarithmic divergence of the specific
heat.

Theoretical treatment
of 2D Ising model with ferromagnetic impurity bonds was pioneered by
Vl.Dotsenko and Vik.Dotsenko (DD) [2-6].
They predicted new critical behaviour of the specific heat \cite{D1,D2},
spin-spin correlation function, magnetization and
magnetic susceptibility \cite{D3,D4}.
Later the same model has been considered in a number of theoretical
works [7-15].
Some authors claimed the specific heat to remain finite for all temperatures
\cite{Ziegler,Timonin}.
The others [7-12]
confirmed DD result for the specific heat, but stated
the critical behaviour of spin-spin correlation function and magnetization
to be the same, or almost the same, as in the pure case,
only slightly changed by logarithmic corrections.

Experiments on quasi--2D compounds with almost Ising spins [16-20] did not show
any deviations from Onsager results, probably because of unsufficient
accuracy, caused by large scale inhomogeneities, which smooth out
the phase transition.
Moreover, in such experiments it is not
possible to exclude 3D effects. So, to understand, if any of the theories
catch essential physics, it is necessary to perform computer simulations.

Early Monte Carlo simulations \cite{MC1,MC2} demonstrated the same
critical behaviour as in the pure case.

The large-scale simulations were started by Andreichenko et al [23-25].
The specific
heat at the critical point of the model with impurities $C(T_c)$ was studied
as a function of the system size $L$ \cite{A2,A3}. For large impurity
strength the specific heat $C(T_c)$ was found to be
proportional to $\log \log (L)$, which
seems to be in agreement with DD prediction.

The behaviour of magnetization $M(T_c)$ and magnetic
susceptibility
$\chi (T_c)$ as a function of $L$ turned out \cite{A2,A3} to be the same
as in the pure case, which contradicted to DD theory.

Usually critical behaviour is studied as a function of relative temperature
distance
$
\tau = (T_c-T)/T_c
$
from  $T_c$.
Some deviations from the pure behaviour of $M(\tau)$ and $\chi (\tau)$
were found by Wang et al \cite{A3}, and in
Special Purpose Processor (SPP) simulations \cite{SPC1,SPC2}.
But the accuracy of [25-27] was not very high,
and the critical region was not defined.

The new SPP \cite{NSPC1,NSPC2}, using cluster Wolff algorithm,
permits to get very accurate
description of the 2D random Ising model critical behaviour.
It gives us not only
the thermodynamic quantities, but also the spin-spin correlation function
\cite{NSPC2,corpaper}
$$
C\!F(r) = <S(0)S(r)> \; ,
$$
which is directly studied by theories [4-12].

\section{The model}

We study the same system which has been considered earlier [25-27].
Ising spins are located at the nodes of two-dimensional square lattice.
To avoid appearance of the border-induced terms, we use periodic
boundary conditions in both directions.

Disorder is introduced by random distribution of impurity coupling constants
over lattice bonds.
Pure case exchange interaction constant is denoted as $J$, and impurity
coupling constant is $J'$. The probability to find on some bond $J'$ is $p$,
and the probability to find $J$ is $(1-p)$.
In this paper we do not consider spin glasses, so both $J$ and $J'$ are
ferromagnetic.

What can be said about the phase diagram of such a model ?
At $p=0$ there is the Onsager phase transition at
$(1/T_c) \approx 0.44068$, if the pure coupling constant $J$
is chosen to be equal to $1$.
The shift of the $T_c$ due to impurities for small $p$
was calculated exactly in \cite{Harris,Tsallis} for $J'=0$, and in
\cite{D2} for the general case.

Certainly, impurities destroy thermal fluctuations and decrease $T_c$.
If there are enough of strong impurities, then the phase transition
disappears. Indeed, if $J'=0$, and $p>0.5$, then there is no
percolation of bound spins in the system \cite{Sykes}.
The lattice is broken into
finite islands of interacting spins, and in a finite system there can be
no phase transition.

Fortunately, for $p=0.5$ and $J' > 0$
there exist exact duality relation \cite{Fisch,Derrida},
which gives the value of $T_c$ as a function of $J$ and $J'$:
\begin{equation}
\tanh (J/T_c) = \exp( -2 J'/T_c)\; .     \label{dual}
\end{equation}
In \cite{A1,A3} it was checked, that indeed, the position of the
specific heat maximum tends to this $T_c$ in the limit
$L \rightarrow \infty$.
This confirms that there is only one phase transition point in the system.

In their theory of the impure Ising model DD \cite{D5} introduced a small
parameter
$$
g \propto p (J-J')^2
$$
and the impurity induced length $l_i$
$$
\log (l_i) \propto \frac{1}{g} \; .
$$
The influence of impurities should be important only
on the distances larger than $l_i$.
If the disorder is small, $g \ll 1$, then for the finite lattice
with the linear size $L$ it may happen that $l_i \gg L$. In this case
it is impossible to notice the influence of impurities on the
critical behaviour.

For this reason to study the deviations caused by inhomogeneities,
we must choose
$g$ as large as possible to decrease $l_i$. So, in real MC simulations
$g$ cannot be very small, and exact theoretical
formula for $g$ by DD \cite{D5}, is no more valid. Nevertheless,
as we show later, the DD formula for the specific heat
\begin{equation}
C(T) \propto \frac {1}{g} \log(1 + g \log (\frac{1}{\tau}))\; ,  \label{C1}
\end{equation}
where $g$ is regarded just as
a parameter, reasonably describes the simulation data.

\section{Simulation procedure}

We used for the simulations the first cluster algorithm SPP \cite{NSPC1,NSPC2}.
It implements in hardware very efficient cluster Wolff algorithm [35-37],
which is the improved version of the Swendsen - Wang algorithm \cite{SW}.
The Wolff algorithm does not suffer of the critical slowing down,
and at the critical point it should be about $L^2$ times faster than
the conventional one spin-flip algorithm.

 In the case of our SPP $L$ can be as large as 1024.
So, the improvement in speed is about one million times.

Detailed description
of the SPP structure and functioning can be found in \cite{NSPC2}. The class of
problems, which can be solved by the SPP, is described in \cite{NSPC1,NSPC2}.

Precise meaning of the critical slowing down absence was found
in the simulations of the pure case
\cite{NSPC2} :
the relaxation time, measured in real simulation time, is the same
at the critical point and far from it.

The relaxation time is defined
in the following way: we start simulation with all spins pointing
in one direction. After some time all the thermodynamic values,
such as magnetization or the lattice energy, come into the zone of
thermal fluctuations near thermal equilibrium. We call this time
the relaxation time.

For the pure case it is necessary to flip about 20 Wolff
clusters to get to the fluctuation region near $T_c$, and about
5 Wolff clusters far from $T_c$ for $L=1024$ \cite{NSPC2}. On the other hand,
mean number of spins in the cluster far from the critical point
is almost equal to the total number of spins $L^2$, and near $T_c$
the number of spins in the cluster is about 4 times lower.
So, the relaxation time, measured as the really spend computer time,
is the same.

The situation in the disordered system can be seen on Fig.1 and 2.
All results are given for the case $J=1$, $J'=0.25$, $p=0.5$.
According to (\ref{dual}) this corresponds to $(1/T_c) = 0.80705186$.

Fig.1 shows the relaxation of magnetization
and the correlation function $C\!F(L/2)$ near $T_c$, at $(1/T_c) = 0.808$.
We see that again, like in the pure case, 20 clusters should be flipped
to enter the thermal fluctuation zone near $T_c$.

As can be seen from Fig.1, there is some correlation between
$M$ and $C\!F(L/2)$.

Indeed, there are two ways to define $M$.
First, we can count the difference in the number of up
$N_{\uparrow}$ and down $N_{\downarrow}$
spins in the system, then the magnetization is given by
$$
M_1 = (N_{\uparrow}-N_{\downarrow})/(N_{\uparrow}+N_{\downarrow})\; .
$$
The second definition of $M$ is usually used to find it theoretically
for the infinite system:
$$
M = (<S(0)S(\infty)>)^{1/2}\; .
$$
In the finite system we can alternatively define $M$ as
$$
M_{2} = (C\!F(L/2))^{1/2}\; .
$$
Our simulations show that $M_{1}$ and $M_{2}$,
averaged over impurity distribution,
normally lead to the
same mean values of magnetization. Noticeable difference between them
appears only very close to $T_c$, when the finite lattice size
effects come into play. Nevertheless, the correlation between
$M$ and $C\!F(L/2)$ persists up to $T_c$, as can be seen
from Fig.1.

The relaxation time is about 20 clusters not only for $M$, but
also for other thermodynamic quantities.
That can be seen
in Fig.2,
which describes the relaxation of the neighbour
spins correlation function $C\!F(1)$.
In the pure case $C\!F(1)$ coincides with the energy per one bond.
Now it is not so, because the energy is given by the
mean value of
$$
<J_{01}S(0)S(1)>
$$
and there are some correlations between the value of coupling
constant $J_{01}$ on the bond, connecting two neighbour nodes $0$ and $1$,
and the sign of the two spin product.
Nevertheless, the relaxation curve for the energy, also shown in Fig.2,
behaves very much like relaxation curve for $C\!F(1)$.

Fluctuations of the magnetization, shown on Fig.1, are very large,
as should be the fluctuations of the order parameter near the
critical point.
On the other hand, the energy, which is not the
order parameter, does not fluctuate so strongly.

Because the relaxation time near $T_c$ in the impure case
is the same as in the pure case, we again come to the conclusion
that the critical slowing down is absent.

Nevertheless, to get all the thermodynamic data, described below,
we permitted spins of the each sample to relax during first
2000 cluster flips. Only after that measurements for each sample
were started.

Each sample has its own distribution of impurity bonds. Coupling
strengths $J=1$ and $J'=0.25$ were ascribed to each bond with
the probability one half to insure selfduality condition.
It is the difference between different samples which determines
the value of standard errors for all the thermodynamic data.

We use 3 sets of data. Low accuracy data, obtained with 10
samples, describe large $\tau$ region. Better data, obtained
with 100 samples, give general picture for all temperatures.
Finally, very accurate 1000 samples data were obtained in the
critical region, where asymptotics can be defined.


The importance of the proper determination of the critical region
can be seen from Fig.3 and 4,
on which we show reduced magnetization
and magnetic susceptibility for the pure case, $L=1024$ \cite{NSPC2}.
In the pure case the critical behaviour is described by power
laws
\begin{equation}
M_0 = 1.22241\, (\tau)^{1/8}            \label{m0}
\end{equation}
\begin{equation}
\chi_0 =  (0.025537-0.001989\, \tau')\, (\tau')^{-7/4}/T   \label{h0}
\end{equation}
where $\tau' = (T_c - T)/T $  \cite{Wu}.

Fig.3 and 4 show ratios of the pure case \cite{NSPC2} $M$ and $\chi$
to this asymptotic laws.
It is clear, that the critical region, in which the asymptotics
are valid, is not very wide:
\begin{equation}
0.001 < \tau <0.02                                        \label{pregion}
\end{equation}
Low $\tau$ restriction comes from finite lattice effects, which
become important for
\begin{equation}
\tau < (1/L) \; .                                       \label{llimit}
\end{equation}
On the other hand, we know
the exact solution for the magnetization of the infinite system
$$
M_{\infty} = (1 - \frac{1}{\sinh^{4}(2/T)})^{1/8}\; .
$$
This solution is true at large $\tau$, and shows that there should be
analytic corrections to simple scaling law (\ref{m0}). These
corrections to scaling lead to the large $\tau$ limit in (\ref{pregion}).
\section{Results}

In the impure case asymptotics (\ref{m0},\ref{h0}) are no longer valid.
Nevertheless, deviations from them are not very large. To see these
deviations better, it is convenient again to divide $M$ and $\chi$
by $M_0$ and $\chi_0$.
Corresponding ratios are shown as a function of $\tau$ on Fig.5 and 6.

{}From this figures it is obvious, that the critical behaviour is changed,
and can be described by larger effective exponents than in the pure case.
It is also obvious, that the critical region for the impure case
is somewhere in the limits
\begin{equation}
0.003 < \tau  < 0.03  \; .                          \label{iregion}
\end{equation}
To check once more, that small $\tau$ behaviour is determined by
the finite lattice size, we show the change of this behaviour with
$L$ on Fig.7 and 8.
{}From Fig.8 we see that the maximum of $\chi$
is shifted from $\tau =0.003$ for $L=1024$ to $\tau =0.006$ for $L=512$,
as can be expected.

It is natural to investigate critical region (\ref{iregion}) more carefully.
The $M$ and $\chi$ results, obtained in this region using 1000 samples
for L=1024, are shown in Fig.9 and 10.
They can be described by changed effective exponents in power
laws (\ref{m0},\ref{h0}).
This does not contradict to the description in terms of
logarithmic corrections to that power laws \cite{A3}, but requires less
number of fitting parameters. The effective exponents should not be
regarded as real exponents, because that will violate the
standard scaling relations.

Fig.11
shows the same data, as in Fig.9, additionally divided by
$\tau ^{\epsilon}$, for $\epsilon = 0.009$, $0.0075$, $0.006$.
We see, that the effective exponent of $M$ in the critical
region is increased by $0.0075$ from the pure Ising value $0.125$.

Fig.12
shows the analogues data for the magnetic susceptibility $\chi$.
In this case $\epsilon = -0.11$, $-.135$, $-.17$. This implies
that the effective critical exponent for $\chi$ is $1.75+0.135$.


The specific heat $C(\tau)$ is more difficult to study than the
magnetization for two reasons.

1) The specific heat is obtained as fluctuations of the energy
according to the formula
$$
C = < (E-<E>)^2 >/T^2 \; .
$$
As a result, fluctuations of $C(\tau)$ for a given $\tau$ do not
decrease with the increasing lattice size $L$, as do the fluctuations
of $M$. In reality, standard deviations of $C(\tau)$ depend only on $\tau$
and the number of flipped clusters, which was used to measure $C(\tau)$,
but not on $L$.

For this reason it has a sense to get large $\tau$
values of $C(\tau)$,
for which the influence of finite lattice size is not important,
on smaller lattices, which requires less computer time.

2) The specific heat $\tau$ dependence is complicated. This
creates difficulties in defining the critical region.

Despite these two problems, the DD formula (\ref{C1})
can help us to interpret the simulation data.

Fig.13
describes general behaviour of the specific heat in both
pure and impure cases as a function of $\tau$. We see, that
impurities reduce $C(\tau)$ and cause deviations
from the simple $\log (\tau) $ asymptotic behaviour.

It is convenient to study the difference between $C(\tau)$ and supposed
asymptotics of it.

Pure case asymptotic of $C(\tau)$ per one node for
$\tau \rightarrow 0$ is
\begin{equation}
C_0^{pure}(\tau) = 0.4945 \log(\frac{1}{\tau})    \label{c0}
\end{equation}
plus some constant. Analytic "corrections to scaling" can be made visible,
if we draw the difference $y_{pure}(\tau)$ between the exact solution for the
infinite lattice $C_{exact}^{pure}(\tau)$ and (\ref{c0}). This difference
\begin{equation}
y_{pure}(\tau) = C_{exact}^{pure}(\tau) - 0.4945 \log (\frac{1}{\tau}) + 0.6
\end{equation}
is shown in Fig.14.
Small deviations from horizontal line at large $\tau$
demonstrate "corrections to scaling" for the simple
logarithmic behaviour of $C_{exact}^{pure}(\tau)$.

The impure data in the same Fig.14 are described by the curve
\begin{equation}
y(\tau) = C(\tau) - 0.21 \log (\frac{1}{\tau})
\end{equation}
The coefficient 0.21 before $\log (\tau)$ in this formula was
chosen to make $y(\tau)$ curve horizontal near $\tau = 0.01$.
We see, that deviations from the pure $\log$ behaviour are
larger in the impure case.

For this reason we may try to approximate $C(\tau)$ by (\ref{C1}).
It is natural to try to choose proportionality coefficient in such a way
that for $g \rightarrow 0$ this formula turns to the pure Ising
formula. Then $C(\tau)$ must be chosen as
\begin{equation}
C(\tau) = \frac {0.4945}{g} \log(1 + g \log (\frac {1}{\tau})) +const\; ,
\label{C2}
\end{equation}
where $g$ and $const$ are parameters to be found from comparison
with the simulation data.

Fig.15
shows
\begin{equation}
z(\tau)=C(\tau)-\frac{0.4945}{0.295}\log(1+0.295 \log(\frac{1}{\tau}))+0.6\; .
\end{equation}
Here $g=0.295$ was chosen to make the curve as horizontal as possible
in the critical region $\tau > 0.003$.
Comparison with the pure case curve on the same Fig.15 shows that
"corrections to scaling" for the DD formula (\ref{C2}) are approximately
of the same value as in the pure case.

The choice of $g = 0.295$ is justified by Fig.16,
which shows 1000 samples data for $L=1024$ in the critical region.
This Fig.16 gives curves $z(\tau)$ for 3 different values
of $g$ : $0.31$, $0.295$, $0.28$.

We can conclude that DD formula (\ref{C2}) is surprisingly good in description
of the impure case specific heat. The value of $g$ is in a reasonable
agreement with that found from
logarithmic correction fit of magnetization data in \cite{A3}.

\section{Conclusions}

More pronounced singularities
of magnetic properties $M(\tau)$ and $\chi(\tau)$ in the impure case
are in qualitative agreement with the statement \cite{Shalaev2} that
$\chi$ and $M$ should remain the same functions of the
correlation length $\xi$, as in the pure case
\begin{equation}
\chi \propto \xi^{7/4}   \; ,            \label{chi}
\end{equation}
\begin{equation}
M \propto \frac{1}{\xi^{1/8}}   \; ,     \label{M}
\end{equation}
but $\xi$ grows faster than $(1/ \tau)$ when $\tau \rightarrow 0$.

So, the impurities increase the correlation length $\xi(\tau)$
for a given $\tau$. At a first glance, that seems to contradict
to common sense. But we should take into account, that the impurities
decrease the critical temperature $T_c$, so the increased $\xi(\tau)$
is obtained at much lower absolute temperature $T$.

It would be natural to say that our data give increased values of
magnetization critical exponent $\beta$ and magnetic susceptibility
critical exponent $\gamma$. But this would violate the standard
scaling formula $\alpha + 2\beta + \gamma=2$. For this reason we regard
increased $\beta$ and $\gamma$ as effective exponents. This effective
exponents can be treated as describing logarithmic corrections
to the correlation length $\xi \approx (1-g\log(\tau))^{1/2}/\tau$
\cite{D1}, combined with large analytic in $\tau$ corrections
to the scaling laws (\ref{chi},\ref{M}).\\[2.5ex]


{\bf Acknowledgments}
\\[1.5ex]
We are grateful to VL. S. Dotsenko who was the initiator of all
recent Monte Carlo simulations of the random Ising Model.
We also would like to thank V. L. Pokrovsky, W. Selke, A. Compagner,
H. W. J. Blote, J. R. Herringa, A. Hoogland and I. T. J. C. Fonk
for many helpful discussions.

This work is partially supported by grants 93-02-2018 of RFFR,
the Russian Foundation for Fundamental Research and 07-13-210 of NWO,
the Dutch Organization of Scientific Research.


\newpage

{\bf Figure captions}\\[1ex]

\noindent
{\bf Fig.1:} Time relaxation of the magnetization $M$ (solid triangles)
and of the correlation function $C\!F(L/2)$ (empty triangles) as
a function of a number $Nc$ of flipped Wolff clusters. Broken lines
show mean values, obtained for a large number of flipped clusters
for the linear lattice size $L=1024$.
Temperature $T$ is very close to $T_c$.\\[1ex]

\noindent
{\bf Fig.2:} Time relaxation of the neighbour spin-spin correlation
function $C\!F(1)$ and of the energy per one bond (insert). Broken
lines have the same meaning as in Fig.1.\\[1ex]

\noindent
{\bf Fig.3:} Ratio of the magnetization $M$, obtained by the cluster
SPP \cite{NSPC2} for the pure Ising model, $L=1024$, to the
asymptotic law $M_0(\tau)$.\\[1ex]

\noindent
{\bf Fig.4:} Ratio of the magnetic susceptibility $\chi$, obtained by
the cluster SPP \cite{NSPC2} for the pure Ising model, $L=1024$, to the
asymptotic law $\chi_0(\tau')$.\\[1ex]

\noindent
{\bf Fig.5:} Ratio of the random bond Ising Model magnetization to the
pure case asymptotic law $M_0(\tau)$. Down triangles show
10 samples data, up triangles -- 100 samples data.\\[1ex]

\noindent
{\bf Fig.6:} Ratio of the random bond Ising Model magnetic susceptibility
to the pure case asymptotic law $\chi_0(\tau')$. Down triangles show
10 samples data, up triangles -- 100 samples data.\\[1ex]

\noindent
{\bf Fig.7:} Impure case $M(\tau)/M_0(\tau)$ for two different
lattice sizes. Empty circles show $L=512$ data, solid
diamonds -- $L=1024$ data.\\[1ex]

\noindent
{\bf Fig.8:} Impure case $\chi(\tau')/\chi_0(\tau')$ for
$L=512$ (empty circles) and $L=1024$ (solid diamonds).\\[1ex]

\noindent
{\bf Fig.9:} 1000 samples critical region data for the
impure case $M(\tau)/M_0(\tau)$, $L=1024$.\\[1ex]

\noindent
{\bf Fig.10:} 1000 samples critical region data for the
impure case $\chi(\tau')/\chi_0(\tau')$, $L=1024$.\\[1ex]

\noindent
{\bf Fig.11:} Impure case $M/(M_0 \tau^{\epsilon})$ for
$\epsilon=0.006$ (down triangles), $0.0075$ (diamonds),
$0.009$ (up triangles).\\[1ex]

\noindent
{\bf Fig.12:} Impure case $\chi/(\chi_0 (\tau')^{\epsilon})$ for
$\epsilon= -0.11$ (down triangles), $-0.135$ (diamonds),
$-0.17$ (up triangles).\\[1ex]

\noindent
{\bf Fig.13:} Specific heat per one node $C(\tau)$ for the $L=1024$
impure system (up triangles), $L=512$ impure system (empty boxes),
and infinite pure Ising system (broken line).\\[1ex]

\noindent
{\bf Fig.14:} Difference $y(\tau)$ between $C(\tau)$ and the best logarithmic
approximation for it. Pure case $y_{pure}(\tau)$ is shown by the broken line.
Up triangles describe $L=1024$ impure data, empty boxes -- $L=512$ impure
data.
Number of samples is equal to 100 in both cases.
Only the dependence or independence of $y$ on $\tau$ is important,
so all the differences in Fig.13-Fig.16 are displaced by arbitrary
constant.\\[1ex]

\noindent
{\bf Fig.15:} Difference $z(\tau)$ between $C(\tau)$ and the DD
approximation for $g=0.295$. Up triangles describe $L=1024$ data,
empty boxes -- $L=512$ data. Broken line shows infinite pure system
$C(\tau)$ deviations from the simple $\log(\tau)$ low. This deviations
are caused by analytic corrections to scaling. Decrease of $z(\tau)$
for $\tau < 2 10^{-3}$ is caused by finite lattice effects.\\[1ex]

\noindent
{\bf Fig.16:} 1000 samples, $L=1024$,
critical region difference $z(\tau)$ between $C(\tau)$
and DD approximations for $g=0.28$ (down triangles), $0.295$ (diamonds),
$0.31$ (up triangles)

\end{document}